\documentclass[12pt]{article}

\usepackage[T1]{fontenc}

\usepackage{amsfonts,mathtools, amsmath,amsthm,amssymb,mhequ,dsfont}

\usepackage{graphicx,booktabs,xcolor,rotating,float}

\usepackage[top = 2cm , bottom= 2cm , left= 2cm , right= 2cm]{geometry}
\usepackage[colorlinks,citecolor=gr,linkcolor=gr,urlcolor=gr]{hyperref}

\definecolor{prp}{HTML}{b16286}
\definecolor{bl}{HTML}{4E09B2}
\definecolor{gr}{HTML}{61635F}

\usepackage{titling}
\usepackage{setspace}
\usepackage{authblk}

\usepackage{titlesec}
\titleformat{\section}
  {\centering\bfseries}
  {\thesection}
  {.5em}
  {\MakeUppercase}

\titleformat{\subsection}
  {\normalfont\bfseries\itshape}{\thesubsection}{1em}{}

\usepackage[shortlabels]{enumitem}
\usepackage[font=small,labelfont=bf]{caption}

\usepackage{mdframed}
\mdfdefinestyle{st}{%
    linecolor=bl,
   linewidth=2pt,
        }

\usepackage[ruled,norelsize,vlined]{algorithm2e}
\usepackage[noabbrev,capitalise]{cleveref}
 \AtBeginDocument{%
\Crefname{algocf}{algorithm}{algorithms}
    \Crefname{section}{section}{sections}%
    \Crefname{figure}{Figure}{Figures}%
}

\usepackage{natbib}

\newcommand{\w}{\mathbf{w}}
\newcommand{\bEta}{\boldsymbol{\eta}}
\newcommand{\wt}{\mathbf{w}^{\intercal}}
\newcommand{\Wni}{{\mathbf{W}}_{-i}}
\newcommand{\Wi}{{\mathbf{w}}_{-i}}
\newcommand{\W}{\mathbf{W}}

\newcommand{\z}{\mathbf{z}}
\newcommand{\E}{\mathbb{E}}
\newcommand{\I}{\mathbf{I}}
\newcommand{\bS}{ \mathbf{S} }
\newcommand{\ba}{ \mathbf{a} }

\newcommand{\bSigma}{ {\boldsymbol \Sigma} }

\newcommand{\bmu}{ {\boldsymbol \mu} }

\newcommand{\blambda}{ {\boldsymbol \lambda} }

\begin{document}

  \title{Stratified stochastic variational inference for high-dimensional network factor model}
\author[1]{Emanuele Aliverti}
\affil[1]{Department of Economics, University Ca' Foscari Venezia}

\author[2]{Massimiliano Russo}
\affil[2]{Harvard--MIT Center for Regulatory Science, Harvard Medical School and Department of Data Science Dana-Farber Cancer Institute}
\date{}
\maketitle
\begin{abstract}
	\noindent
There has been considerable recent interest in Bayesian modeling of high-dimensional networks via latent space approaches. When the number of nodes increases, estimation based on Markov Chain Monte Carlo can be extremely slow and show poor mixing, thereby motivating research on alternative algorithms that scale well in high-dimensional settings. In this article, we focus on the latent factor model, a widely used approach for latent space modeling of network data. We develop scalable algorithms to conduct approximate Bayesian inference via stochastic optimization. Leveraging sparse representations of network data, the proposed algorithms show massive computational and storage benefits, and allow to conduct inference in settings with thousands of nodes. An \textsc{r} package with an efficient \textsc{c++} implementation of the proposed algorithms is provided.
\end{abstract}

\noindent%

\vfill

\section{Introduction}
\label{sec:intro}
Network data are routinely collected and analyzed in different fields of science; for example, neuroscience \citep{bullmore:2009}, genetics \citep{wu:2008} and epidemiology \citep{keeling:2005}, among many others. Refer also to \cite{newman:2018} for an introduction to network data and their analysis.
One of the main goals in network data analysis is to characterize the geometry underlying node relationships, providing a parsimonious, yet flexible, representation of the connectivity patterns. 
This goal can be achieved modeling the connectivity architectures in terms of a low-dimensional latent structure, where the edges are represented as conditionally independent random variables given a set of latent coordinates \citep[e.g.,][]{hoff:2002}.
Beside improving computation, these approaches provide concrete benefits in interpretation; for example, the latent structure can be related with observable covariates to improve the understanding of connectivity patterns, or divided into clusters to detect nodes that behave similarly in terms of their unobservable features \citep[e.g.,][]{aliverti:2019}. 
The increasing availability of network data has further motivated the development of novel latent structure models for networks, covering more complex settings such as dynamic networks \citep[e.g.,][]{durante:2014,sewell:2015}, multilayer networks \citep[e.g.,][]{gollini:2016,dangelo:2019} and populations of networks \citep{durante:2017}.
Since the number of edges grows quadratically with the number of nodes, representing the latent structure of a network is computationally challenging, even for networks with few hundred nodes.
This issue stimulates the development of novel methods and computational routines to accommodate large network structures efficiently, leveraging different network properties---such as sparsity or block structures---to facilitate computations.
For example, large network data are often very sparse, with the number of observed edges being much smaller than the total number of possible connections. 
This feature allows to parsimoniously store network data via sparse matrix representations, or, equivalently, edge-lists formats \citep[e.g.,][]{igraph}.
However, estimation of latent structure models often requires edge-specific operations, lowering the benefits of these representations; see for example \citet{ho:2016} for related arguments.
In addition, most inference procedures for latent space models are based on Markov Chain Monte Carlo (\textsc{mcmc}) algorithms \citep[e.g.,][]{hoff:2018}, and these methods scale poorly with the number of nodes.

There have been some attempts to improve computational efficiency and mixing of \textsc{mcmc} for latent space model for networks.
For example, \citet{raftery:2012} derive an unbiased estimator of the log-likelihood based on an informative subset of nodes, and successfully perform \textsc{mcmc}  estimation in an application with roughly three thousands nodes \citep[Section 4.2]{raftery:2012}.
Although  this approach effectively reduces the cost of each likelihood evaluation, inference is still computationally demanding; the algorithm requires a preliminary pilot \textsc{mcmc} run and a considerable amount of storage to perform Monte Carlo integration.

These computational issues motivate the development of scalable methods for approximate Bayesian inference, with Variational Bayes (\textsc{vb}) being a popular option. 
Specifically, \textsc{vb} algorithms approximate the posterior distribution via optimization, estimating the closest member (in Kullbak-Lielber divergence) within a pre-specified class of distributions.
This class includes some restrictions to achieve computational tractability; for example, the resulting approximate posterior distribution factorizes in independent blocks of parameters (mean-field \textsc{vb}), or follows a specific parametric form (e.g., multivariate Gaussian). 
Refer to \cite{blei:2017} and references therein for more details. 
To date, the currently available \textsc{vb} routines for latent space models \citep{gollini:2016,vblcp} are based on restrictive assumptions on the functional form of the variational distributions and rely on several approximations of the complete log-likelihood function \citep{gollini:2016}. In addition, these algorithms encounter computational issues in high-dimensional settings, since they scale quadratically with the number of nodes \citep{vblcp}.

In this paper, we focus on the Latent Factor Model \citep[\textsc{lfm},][]{hoff:2005}---also referred to as random dot-model \citep{young:2007}---for binary undirected networks, a very general approach that can approximate several latent position models under mild conditions \citep{tang:2013,athereya:2018}.
We propose scalable algorithms based on a stratified stochastic variational approximation for the \textsc{lfm}, which we refer to as \textsc{svilf} in the following.
From a computational perspective, \textsc{svilf} explicitly relies on sparse network representations via edge-lists, which allows to scale up computations for networks consisting of thousands of nodes.
Leveraging a conditionally conjugate exponential family representation, we provide a unified framework for the \textsc{lfm} with logistic and probit link function and illustrate how \textsc{svilf} can be directly implemented under both specifications.

\section{Methods}
\subsection{Latent Factor Model}
Let $\mathbf{Y}$ denote an $n \times n$ binary symmetric adjacency matrix with elements $y_{ij} = y_{ji} = 1$ denoting the presence of an edge between node $i$ and node $j$ with $i=2,\dots,n$ and $j=1,\dots,i-1$, and $y_{ij} = y_{ji} = 0$ otherwise.
The \textsc{lfm} for networks parametrizes the edges as conditional independent Bernoulli random variables given a set of latent positions ${\w_i = (w_{i1}, \ldots, w_{iH})\in \mathbb{R}^H}$, $i=1,\dots,n$.
This choice implies
\begin{equation}
	\label{eq:lfm}
	(y_{ij} \mid \pi_{ij}) \sim \mbox{Ber}(\pi_{ij}), \quad
	g(\pi_{ij}) = \wt_i\w_j=  \sum_{h=1}^H w_{ih}w_{jh}, \quad i = 2, \dots, n,\, j=1,\dots,i-1,
\end{equation}
where $g: [0,1] \rightarrow  \mathbb{R}$ is a monotone link function.
Popular choices for $g$ include the logit and probit link \citep[e.g.,][]{agresti:glm}. See also \citet{young:2007} for alternative specifications of the model outlined in Equation~\eqref{eq:lfm}.
In a Bayesian setting, we typically assign independent Gaussian priors to the latent factors, thereby letting ${\w_i\sim\mbox{N}_H(\ba_0,\I_H)}$ for $i=1,\ldots,n$. A prior mean $\ba_0 = (a_{01}, \dots, a_{0H})$ different from zero allows to center the factors around an expected network sparsity, accounting for the baseline probability of observing a connection.
According to \Cref{eq:lfm}, the probability of observing a connection between node $i$ and node $j$  depends on their latent positions $\w_i$ and $\w_j$. 
The more similar node $i$ and node $j$ are in the latent space, the more likely it is to observe an edge connecting them.
The similarity between two nodes in the latent space is computed with the multiplicative effect $\w_i^\intercal\w_j$, and this measure is particularly helpful in characterizing transitivity and uncovering group structures \citep{hoff:2018}.

The \textsc{lfm} reduces the number of free parameters needed to characterize the network from $n(n-1)/2$ to $nH$, providing a low-dimensional representation of the connectivity patterns.
Additionally, the \textsc{lfm} is a very flexible latent structure model for networks, since several latent position models can be represented as Equation~\eqref{eq:lfm} when $H$ grows \citep{tang:2013}.
For these reasons, the \textsc{lfm} has been used as a building block for many generalizations involving, among others, time-varying and covariate-dependent networks \citep[e.g.,][]{durante:2014,sewell:2015}.
Therefore, efficient algorithms for the \textsc{lfm} are crucial to analyze large network data, which are routinely collected in different fields of application.

Recalling \Cref{eq:lfm}, it is worth highlighting that the predictor $g(\pi_{ij})$ is linear in the latent factors. 
Focusing on a single factor $\w_i$, we can recast the model outlined in \Cref{eq:lfm} as a conditional binary regression, given the other factors $\w_j$, $j\neq i$.
Denoting with $\mathbf{y}_i = (y_{i1}, \dots,y_{i\,i-1},y_{i\,i+1},\dots, y_{in})$  the $i$-th row of the adjacency matrix $\mathbf{Y}$,
with $\boldsymbol{\pi}_{i}=(\pi_{i1}, \dots,\pi_{i\,i-1},\pi_{i\,i+1},\dots, \pi_{in})$ the $(n-1)$-variate vector of associated edge probabilities and with  $\Wni$ the $(n-1)\times H$ dimensional matrix obtained stacking the factors $\{\w_j\}_{j\neq i}$, it holds that
\begin{equation}
	\label{eq:logw}
	g(\boldsymbol{\pi}_{i}) = \Wni\w_i,
\end{equation}
where the link function $g(\cdot)$ in \Cref{eq:logw} is applied element-wise. 

\Cref{eq:logw} allows one to rely on iterative algorithms for posterior inference under the \textsc{lmf}, considering $n$ binary regressions where each factors acts, in turn, as a regression coefficient. Moreover, for some specific link functions $g$, data-augmentation schemes are available to further simplify posterior inference. 
Specifically, with the logit link function conditional conjugacy can be retrieved relying on the P\`{o}lya-Gamma (\textsc{pg}) data augmentation strategy introduced by \citet{polson:2013}, that leads to the following full conditionals distributions:
\begin{alignat}{4}
	\label{eq:condlogit}
	(z_{ij} \mid \w_i,\w_j, y_{ij}) &\sim \,&& \mbox{\textsc{pg}}(1, \w_i^\intercal\w_j), \quad &&i=2,\dots,n, \quad j=1,\dots,i-1, \\
(\w_i \mid \Wni,\mathbf{z}_i,\mathbf{y}_i) &\sim\, &&\mbox{N}(\boldsymbol{\mu}_{\w_i}, \boldsymbol{\Sigma}_{\w_i}), \quad &&i=1,\dots,n, \nonumber 
\end{alignat}
where
\begin{equation*}
\boldsymbol{\Sigma}_{\w_i}= \left(\Wni^\intercal\mbox{diag}(\mathbf{z}_i)\Wni + \I_H\right)^{-1},\quad \boldsymbol{\mu}_{\w_i}= \boldsymbol{\Sigma}_{\w_i} \left(\Wni^\intercal(\mathbf{y}_i - 0.5\cdot \mathbf{1}_{n-1}) + \I_H\ba_0 \nonumber \right),
\end{equation*}
and $\z_i=(z_{i1}, \dots,z_{i\,i-1},z_{i\,i+1},\dots, z_{in})$. In \Cref{eq:condlogit}, $\textsc{pg}(c,d)$ denotes the density of a P\`{o}lya-Gamma distribution with parameters $c$ and $d$; refer to \citet{polson:2013} for more details.

Similarly, with the probit link function, conditional conjugacy is obtained adapting the data-augmentation proposed in \citet{albert:1993}, which introduces auxiliary observations distributed as truncated normal random variables.
The full conditional distributions under this specification correspond to
\begin{alignat}{4}
	\label{eq:condprobit}
	(z_{ij} \mid \w_i, \w_j, y_{ij}) &\sim&&\begin{cases*} \mbox{\textsc{tn}}(\w_i^\intercal\w_j, 1, [0,+\infty]),\quad  \text{if }\, y_{ij} = 1\\  \mbox{\textsc{tn}}(\w_i^\intercal\w_j, 1, [-\infty,0]),\quad \mbox{if }\,  y_{ij} = 0 \end{cases*}\quad &&i=2,\dots,n, \quad j=1,\dots,i-1,  \\
(\w_i \mid \Wni,\mathbf{z}_i,\mathbf{y}_i) &\sim&&\,\, \mbox{N}_H(\boldsymbol{\mu}_{\w_i}, \boldsymbol{\Sigma}_{\w_i}), \quad &&i=1,\dots,n, \nonumber
\end{alignat}
where
\begin{equation*}
	\boldsymbol{\Sigma}_{\w_i}= \left(\Wni^\intercal\Wni + \I_H\right)^{-1},\quad \boldsymbol{\mu}_{\w_i}= \boldsymbol{\Sigma}_{\w_i}\left(\Wni^\intercal\z_i + \I_H\ba_0\right),
\end{equation*}
and with $\mbox{\textsc{tn}}(\mu,\sigma^2,[a,b])$ denoting a truncated normal distribution with parameters $(\mu,\sigma^2)$, restricted over the interval $[a,b]$.
Posterior inference via \textsc{mcmc} relies on iterative sampling from \Cref{eq:condlogit} or \eqref{eq:condprobit}, constructing a Markov chain which has the joint posterior distribution $p(\W,\z \mid \mathbf{Y})$ as a limiting distribution \citep{gelfand:1990}.
Some factors can be omitted in the expressions above; for example,  ${p(\w_i \mid \Wni,\mathbf{z}_i,\mathbf{y}_i) = p(\w_i \mid \Wni,\mathbf{z}_i)}$ in \Cref{eq:condprobit}. However, this redundant notation helps to unify both algorithms under a general specification, as outlined in the following section.

\subsection{Conditional conjugancy and Variational Inference}
\label{sec:conj}
The conditional distributions for the augmented variables $z_{ij}$ in Equation~\eqref{eq:condlogit} and \eqref{eq:condprobit} can be expressed as
\begin{equation}
\begin{split}
\label{eq:fcz}
p(z_{ij} \mid \w_i,\w_j,y_{ij}) \propto \exp\left\{\eta_{ij}\left(\w_i,\w_j \right)z_{ij} - \kappa\left(\eta_{ij}\left(\w_i,\w_j\right)\right)\right\}p(z_{ij} \mid y_{ij}), \\ \quad i = 2, \dots, n,\,j=1,\dots,i-1,
\end{split}
\end{equation}
with $\kappa(\cdot)$ denoting the log-partition function.
Equation~\eqref{eq:fcz} allows one to express the \textsc{lfm} with logit and probit link with the same exponential family structure. When $g$ is the logit link, ${\eta_{ij}(\w_i,\w_j)  = -0.5(\w_i^\intercal\w_j)^2}$ and $p(z_{ij} \mid y_{ij})$ corresponds to the density of a $\mbox{\textsc{pg}}(0,1)$; when $g$ is the probit link, $\eta_{ij}(\w_i,\w_j) = \w_i^\intercal \w_j$ and $p(z_{ij} \mid y_{ij})$ corresponds to the density of a $\textsc{tn}(0,1,[-\infty,0])$ if $y_{ij} = 0$ and $\textsc{tn}(0,1,[0,\infty])$ if $y_{ij} = 1$.

Conditionally on the observed data, the augmented variables and the factors $\{\w_j\}_{j\neq i}$, the full-conditional distribution of each $\w_i$ corresponds to a multivariate Gaussian density,
\begin{equation}
\begin{split}
	\label{eq:fcw}
	p(\w_i \mid \Wni, \mathbf{z}_i, \mathbf{y}_i) \propto \exp\big\{\bEta_{i1}\left(\Wni,\mathbf{y}_i, \mathbf{z}_i\right)^\intercal \w_i + \mbox{vec}\left(\bEta_{i2}\left(\Wni, \mathbf{z}_i\right)\right)^\intercal \cdot \mbox{vec}\left(\w_i\w_i^\intercal\right) - \\  \kappa\left(\bEta_{i1}\left(\Wni,\mathbf{y}_i, \mathbf{z}_i\right),\bEta_{i2}\left(\Wni, \mathbf{z}_i\right)  \right) \big\}, \quad i = 1,\dots,n,
\end{split}
\end{equation}
where
\begin{equation*}
	\bEta_{i1}\left(\Wni,\mathbf{y}_i, \mathbf{z}_i\right) = \Wni^\intercal(\mathbf{y}_i - 0.5\cdot \mathbf{1}_{n-1}) + \I_H \ba_0, \quad \bEta_{i2}\left(\Wni, \mathbf{z}_i\right) = -\frac{1}{2}\left( \Wni^\intercal\mbox{diag}(\mathbf{z}_i)\Wni + \I_H\right), 
\end{equation*}
with the logit link function, and 
\begin{equation*}
 \bEta_{i1}\left(\Wni,\mathbf{y}_i, \mathbf{z}_i\right) = \Wni^\intercal\z_i + \I_H \ba_0, \quad \bEta_{i2}\left(\Wni, \mathbf{z}_i\right) = -\frac{1}{2}\left( \Wni^\intercal\Wni + \I_H\right), 
\end{equation*}
with the probit link function.

This conditional representation of the \textsc{lfm} facilitates the development of algorithms that exploit maximization strategies for binary regression; for example, \textsc{map} optimization via \textsc{em} algorithm or approximate inference via \textsc{vb} \citep{consonni:2007,duranterigon:2017}.
As discussed in Section~\ref{sec:intro}, the focus of the mean-field \textsc{vb} is on finding an approximation of the posterior distribution $p(\W,\z \mid \mathbf{Y})$ within a restricted class of densities $\mathcal{Q}$, specified as 
\begin{equation}
	\label{eq:mf}
	\mathcal{Q} =\left\{q_{\small\textsc{}}(\W,\z) : q_{\small\textsc{}}(\W,\z)=\prod_{i=1}^n q_{\small\textsc{}}(\w_i; \blambda_i)\prod_{i=2}^n \prod_{j=1}^{n-1}q_{\small\textsc{}}(z_{ij}; \psi_{ij})\right\}.
\end{equation}
Note that each factor $\w_i$ is a function of its own variational parameters $\blambda_i$; similarly, the augmented variables $z_{ij}$ are function of the parameter $\psi_{ij}$.
The optimal \textsc{vb} solution $q^\star_{\small\textsc{}}(\W,\z)$ corresponds to the distribution within $\mathcal{Q}$ that minimizes the Kullback-Leibler (\textsc{kl}) divergence, defined as
\begin{equation}
	\label{eq:kl}
	\mbox{{\footnotesize KL}}\left[q(\W,\z) || p_{\small\textsc{}}(\W,\z \mid \mathbf{Y})\right] = \E_{q(\W,\z)}\left[ \log q_{\small\textsc{}}(\W,\z) \right] -	\E_{q_{\small\textsc{}}(\W,\z)}\left[ \log p(\W,\z \mid \mathbf{Y})\right], \quad q(\W,\z)  \in \mathcal{Q}.
\end{equation}
In practice, \textsc{vb} procedures maximize the related objective function
\begin{equation}
	\label{eq:elbo}
	\mbox{{\footnotesize ELBO}}\left[q_{\small\textsc{}}(\W,\z)\right] = \E_{q_{\small\textsc{}}(\W,\z)}\left[ \log p(\W,\z, \mathbf{Y})\right] - \E_{q_{\small\textsc{}}(\W,\z)}\left[ \log q_{\small\textsc{}}(\W,\z) \right], \quad q(\W,\z),  \in \mathcal{Q},
\end{equation}
which corresponds to the negative \textsc{kl} up to an additive constant not depending on the parameters; see, for example, \citet{blei:2017} and \citet{bishop:2006}.
The mean field assumption in \Cref{eq:mf} and the conditionally conjugate representation facilitate a Coordinate Ascent Variational Inference (\textsc{cavi}) routine to maximize \Cref{eq:elbo}, where each distribution is iteratively optimized with respect to the others in an iterative fashion \citep[e.g.,][]{bishop:2006}. 
The variational distributions composing the optimal solution are in the same family of the full-conditional distributions outlined in \Cref{eq:condlogit,eq:condprobit}, and their parameters can be analytically expressed in terms of variational expectations \citep[e.g.,][]{blei:2017}.
Under our specification,
\begin{equation}
	\label{eq:lamv}
	\blambda_{i1} = \E_{q(\Wi,\mathbf{z}_i)}\left[\bEta_{i1}\left(\Wni,\mathbf{y}_i, \mathbf{z}_i\right)\right], \quad
	\blambda_{i2} = \E_{q(\Wi,\mathbf{z}_i)} \left[\bEta_{i2}\left(\Wni, \mathbf{z}_i \right)\right], \quad i=1,\dots,n
\end{equation}
and 
\begin{equation}
	\label{eq:psi}
	\psi_{ij} = \E_{q(\w_i,\w_j)}\left[\eta_{ij}(\w_i,\w_j)\right], \quad i =2,\dots,n,\, j=1,\dots,i-1.
\end{equation}

Algorithm~\ref{alg:caviL} illustrates the \textsc{cavi} algorithms for the \textsc{lfm} with logit link; refer to the Supplementary Materials for the probit link function.
At each iteration $t$, variational expectations outlined in \Cref{eq:lamv,eq:psi} are taken with respect to the currently optimized distribution $q^{(t-1)}$, iterating until convergence. Note also that at each iteration the \textsc{cavi} provides a monotone sequence of the \textsc{elbo}, and convergence to a local optimum is guaranteed \citep{blei:2017,bishop:2006}. 

\begin{algorithm}[H]
	\small
	\caption{\textsc{cavi} for \textsc{lfm} with logit link.} \label{alg:caviL}
	Initialize $\left\{\bS_1^{(1)}, \dots, \bS_n^{(1)}\right\}$ and  $\left\{\bmu_1^{(1)}, \dots, \bmu_n^{(1)}\right\}$. \\
 \For( ){{\normalfont{$t=2$ until convergence}}}
 {
	 {\bf [1]} $q^{(t)}(\w_i)$ is the density of a $\mbox{N}_H(\bmu_i^{(t)}, \bSigma_i^{(t)})$ with $\bmu_i^{(t)}=[-2\blambda^{(t)}_{i2}]^{-1} \blambda^{(t)}_{i1}$, $\bSigma_i^{(t)}=[-2\blambda^{(t)}_{i2}]^{-1}$ and natural parameters
\begin{eqnarray*}
	\blambda^{(t)}_{i1}= \sum_{j\neq i}\bmu^{(t-1)}_j\left[y_{ij}-0.5\right] + \I_H\ba_0, \quad \blambda^{(t)}_{i2}= -0.5\left(\sum_{j\neq i}\bS_j^{(t-1)} \cdot \bar{z}^{(t-1)}_{ij}  + \I_H\right),
\end{eqnarray*}
for $i =1, \dots, n$. In the above expression, $\bar{z}^{(t-1)}_{ij} = 0.5[\xi_{ij}^{(t-1)}]^{}\mbox{tanh}(0.5 \xi^{(t-1)}_{ij})$, where 
\begin{eqnarray*}
	\bS_j^{(t-1)} = \left[ \bSigma_j^{(t-1)} + \left[\bmu_j^{(t-1)}\right]\left[\bmu_j^{(t-1)}\right]^\intercal \right], \quad \xi_{ij}^{(t-1)} = \left[\mbox{vec}\left(\bS_i^{(t-1)}\right)^\intercal\mbox{vec}\left(\bS_j^{(t-1)}\right)\right]^{\frac{1}{2}}
\end{eqnarray*}
{\bf [2]} $q^{(t)}(z_{ij})$ is the density of a $\textsc{pg}(1, \xi^{(t)}_{ij})$, for $i=2,\dots,n$ and $j=1,\dots,i-1$.

}
{\bf Output} $q^\star(\W,\z)=\prod_{i=1}^n q^\star(\w_i)\prod_{i=2}^n \prod_{j=1}^{n-1}q^{\star}(z_{ij})$. 
\end{algorithm}

\section{Stratified stochastic variational inference}
\label{sec:svilf}
The \textsc{cavi} algorithms introduced in Section~\ref{sec:conj} provide efficient routines to perform approximate Bayesian inference under the \textsc{lfm}.
However, when the number of nodes $n$ is extremely large, computational issues might drastically limit the analysis.
For example, in Step [1] of \Cref{alg:caviL}, the updates of each natural parameter $\blambda_i$ involve summing over $(n-1)$ terms, with $i=1,\dots,n$.
Also, conditional conjugacy simplifies the derivation of analytical results, but it requires to update $n(n-1)/2$ augmented observations even when the dimension $H$ of the latent space is small, therefore exacerbating computational and storage issues.
Although variational routines generally require significantly less iterations than \textsc{mcmc} to reach convergence, the overall complexity of the \textsc{cavi} algorithms is still $\mathcal{O}(n^2)$; therefore, \textsc{cavi} provides a viable solution only in settings where $n$ is moderately large \citep{vblcp}.
We introduce a novel approach that allows to approximate the posterior distribution of the \textsc{lfm} for high-dimensional network-data.

A scalable generalization of classical \textsc{cavi} is provided by Stochastic Variational Inference  \citep[\textsc{svi},][]{hoffman:2013}, where stochastic optimization \citep{robbins:1951} is used to reduce the computation cost of \textsc{vb} routines.
We follow a similar perspective and develop a stochastic \textsc{vb} algorithm specifically tailored for sparse network data.
Following \citet{hoffman:2013}, it is useful to rewrite the \textsc{cavi} routines outlined in Algorithms~\ref{alg:caviL} as the solutions of a system of estimating equations, obtained computing the derivatives of the \textsc{elbo} with respect to the parameters of the variational distributions. 

For each fixed factor $q(\w_i; \blambda_i)$, computing the gradient of the \textsc{elbo} outlined in \Cref{eq:elbo} with respect to $\blambda_i$ and equating to $0$ leads to the following estimating equations.
\begin{equation}
	\label{eq:est}
	\E_{q(\Wi,\mathbf{z}_i)}\left[\bEta_{i1}\left(\Wi,\mathbf{y}_i, \mathbf{z}_i\right)\right] -\blambda_{i1} = 0 , \quad
	 \E_{q(\Wi,\mathbf{z}_1)} \left[\bEta_{i2}\left(\Wi, \mathbf{z}_i \right)\right] - \blambda_{i2} = 0.
\end{equation}
See \citet[Section 2.3]{hoffman:2013} for a formal proof.
Step [1] of \Cref{alg:caviL} is obtained replacing the natural parameters with the quantities outlined in Equation~\eqref{eq:fcw}, expanding expectations with the resulting Gaussian or P\`olya-Gamma or moments, and solving for $\blambda_{i1}$ and $\blambda_{i2}$.

Faster algorithms can be obtained  replacing the gradient in \Cref{eq:est} with a computationally cheaper estimate \citep{robbins:1951}.
In particular, we develop an efficient version of this algorithm based on an informative subset of nodes.
We focus on the update of a generic factor $\w_i$, and denote with ${\mathcal{J}_{i0} = \{j:y_{ij}=0\}}$ the set of indices associated with nodes not connected with node $i$.
Similarly, we define $\mathcal{J}_{i1} = \{j:y_{ij}=1\}$, with $|\mathcal{J}_{i0}| = n_{i0}$ and $|\mathcal{J}_{i1}| = n_{i1} = n-n_{i0}$; note that $n_{i1}$ corresponds to the degree of node $i$.
Adapting \citet{duranterigon:2017}, \Cref{eq:est} can be easily decomposed into the contribution of the $n_{i1}$ nodes connected with $i$ and the remaining as follows.
\begin{equation}
\begin{split}
	\label{eq:gradsplit}
\sum_{j \in \mathcal{J}_{i1}}\E_{q(\w_j)}[\w_j]\left(y_{ij}-0.5\right) + \sum_{j \in \mathcal{J}_{i0}}\E_{q(\w_j)}[\w_j]\left(y_{ij}-0.5\right) + \I_H\ba_0 - \blambda_{i1},\\  
-0.5\left(\sum_{j \in \mathcal{J}_{i1}} \E_{q(\w_{j})}\left[\w_j\w_j^\intercal\right]  \cdot \E_{q(z_{ij})}[z_{ij}] + \sum_{j \in \mathcal{J}_{i0}} \E_{q(\w_{j})}\left[\w_j\w_j^\intercal\right] \cdot \E_{q(z_{ij})}[{z}_{ij}]  + \I_H\right) - \blambda_{i2}.
 \end{split}
 \end{equation}
 Similarly, with the probit link
\begin{equation}
\begin{split}
	\label{eq:gradsplitP}
	\sum_{j \in \mathcal{J}_{i1}}\E_{q(\w_j)}[\w_j]\cdot\E_{q(z_{ij})}[z_{ij}] + \sum_{j \in \mathcal{J}_{i0}}\E_{q(\w_j)}[\w_j]\cdot  \E_{q(z_{ij})}[z_{ij}] + \I_H\ba_0 - \blambda_{i1},\\  
 -0.5\left(\sum_{j \in \mathcal{J}_{i1}} \E_{q(\w_{j})}\left[\w_j\w_j^\intercal\right] + \sum_{j \in \mathcal{J}_{i0}} \E_{q(\w_{j})}\left[\w_j\w_j^\intercal\right]+ \I_H\right) - \blambda_{i2}.
 \end{split}
 \end{equation}
 Our strategy relies on noisy estimates of \Cref{eq:gradsplit,eq:gradsplitP}, constructed using an informative subset of nodes $\mathcal{J}_i^\star = \mathcal{J}_{i1} \cup \mathcal{J}_{i0}^\star$, where $\mathcal{J}_{i0}^\star \subset \mathcal{J}_{i0}$ denotes a sample from $\mathcal{J}_{i0}$, with $|\mathcal{J}_0| = n_{i0}^\star \ll n_{i0}$. 
 Therefore, the update for each factor $\w_i$ relies on the random subset $\mathcal{J}_i^\star$, which consists of all the nodes connected with node $i$ and a smaller subset of not connected nodes.
 This approach implicitly assumes that nodes $j \in \mathcal{J}_{i0}$  not connected with $i$ provide little information about its position in the latent space, and therefore we can estimate the contribution of all these disconnected nodes relying only on few units.

A simple strategy leading to a computationally cheap estimate of \Cref{eq:gradsplit,eq:gradsplitP} is to construct the set $\mathcal{J}^\star_{i0}$ relying on random sampling, where each unit $j \in \mathcal{J}_{i0}$ is included in the sample with probability $n_{i0}^{-1}$.
This choice leads to the following unbiased estimator of \Cref{eq:gradsplit}, denoted as $[B(\blambda_{i1}), B(\blambda_{i2})]$ and corresponding to the discrete random variable taking values

\begin{equs}[e:block]
	\label{eq:estL}
	&B_{\mathcal{J}_i^\star}(\blambda_{i1}) = \sum_{j \in \mathcal{J}_{i1}}\E_{q(\w_j)}[\w_j]\left(y_{ij}-0.5\right) + r_{i0} \sum_{j \in \mathcal{J}_{i0}^\star}\E_{q(\w_j)}[\w_j]\left(y_{ij}-0.5\right) + \I_H\ba_0 - \blambda_{i1},\\  
	&B_{\mathcal{J}_i^\star}(\blambda_{i2}) = -0.5\Big(\sum_{j \in \mathcal{J}_{i1}} \E_{q(\w_{j})}\left[\w_j\w_j^\intercal\right]  \cdot \E_{q(z_{ij})}[z_{ij}] + \\ 
	\multicol{2}{r_{i0} \sum_{j \in \mathcal{J}_{i0}^\star} \E_{q(\w_{j})}\left[\w_j\w_j^\intercal\right] \cdot \E_{q(z_{ij})}[{z}_{ij}]  + \I_H\Big) - \blambda_{i2},}
\end{equs}

for each possible $\mathcal{J}_{i0}^\star \subset \mathcal{J}_{i0}$, with $r_{i0} = n_{i0} / n_{i0}^\star$.
Similarly, for \Cref{eq:gradsplitP}
\begin{equs}[e:block]
	\label{eq:estP}
	&B_{\mathcal{J}_i^\star}(\blambda_{i1}) = 	\sum_{j \in \mathcal{J}_{i1}}\E_{q(\w_j)}[\w_j]\cdot\E_{q(z_{ij})}[z_{ij}] + r_{i0} \sum_{j \in \mathcal{J}_{i0}^\star}\E_{q(\w_j)}[\w_j]\cdot  \E_{q(z_{ij})}[z_{ij}] + \I_H\ba_0 - \blambda_{i1},\\  
	&B_{\mathcal{J}_i^\star}(\blambda_{i2}) =  -0.5\left(\sum_{j \in \mathcal{J}_{i1}} \E_{q(\w_{j})}\left[\w_j\w_j^\intercal\right] + r_{i0} \sum_{j \in \mathcal{J}^\star_{i0}} \E_{q(\w_{j})}\left[\w_j\w_j^\intercal\right]+ \I_H\right) - \blambda_{i2}.
\end{equs}

From a storage perspective, this representation allows to explicitly rely on the edge-list format of the network, where only the distinct pairs $(i,j)$ associated with an edge are stored in memory.
In its original formulation, \textsc{svi} relies on sampling one single observation per iteration, and updating the gradient with the resulting estimate.
Instead, for \textsc{svilf} we recommend letting $n_{i0}^\star = \min(n_{i0}, \lfloor \gamma n_{i1} \rfloor)$, with $\gamma > 1$ and where $\lfloor \gamma n_{i1}\rfloor$ denotes the greatest integer less than or equal to $\gamma n_{i1}$.
In sparse network settings, $n_{i0}^\star \gg n_{i1}^\star$ and the random subset $\mathcal{J}_{i0}^\star$ will generally consist of $n_{i0}^\star = \lfloor \gamma n_{i1} \rfloor$ elements.
Larger values of $\gamma$ leads to estimators $[B(\blambda_{i1}), B(\blambda_{i2})]$ with smaller variance, while small values of $\gamma$ lead to more efficient computation.
In our experience, letting $\gamma \in [1,5]$ provides good performance in a large number of applications; see \Cref{sec:sim,sec:app}.

The simple form of the proposed estimator allows the direct applications of the stochastic approximation method proposed in \citet{robbins:1951} to solve \Cref{eq:est} via iterative updates as 
\begin{equation}
	\label{eq:updates}
\begin{split}
	\blambda_{i1}^{(t)} = 	(1-\rho_t) \blambda_{i1}^{(t-1)} + \rho_tB_{\mathcal{J}_t^\star}(\blambda_{i1}^{(t-1)}),\\
	\blambda_{i2}^{(t)} = 	(1-\rho_t) \blambda_{i2}^{(t-1)} + \rho_tB_{\mathcal{J}_t^\star}(\blambda^{(t-1)}_{i2}),
\end{split}
\end{equation}
where	$[B_{\mathcal{J}_t^\star}(\blambda^{(t-1)}_{i1}), B_{\mathcal{J}_t^\star}(\blambda^{(t-1)}_{i2})]$ denotes a draw from the estimators outlined in Equation~\eqref{eq:estL} and \eqref{eq:estP} evaluated at $(\blambda_{i1}^{(t-1)}, \blambda_{i2}^{(t-1)})$, and $\rho_t$ denotes a sequence of step size such that $\sum_t \rho_t = +\infty$ and $\sum_t \rho_t^2 < +\infty$ \citep{robbins:1951}.
A standard choice is to let $\rho_t = (t + \alpha)^{-\beta}$, with $\alpha>0$ and $0.5 < \beta <1$ \citep{hoffman:2013,duranterigon:2017}.
Recalling Appendix A of \citet{hoffman:2013}, in Equation~\eqref{eq:updates} parameters $\blambda_i$ are simultaneously updated to guarantee converge to the solution of \eqref{eq:est}.
These optima are analytically obtained for each $q(z_{ij})$, and iteratively for $q(\w_i)$, conditioning on the values $\{\w_j\}_{j\neq i}$ at the previous iteration, with $i=1,\dots,n$.

As discussed in \citet{raftery:2012} within a different context, uniform subsampling of disconnected nodes might lead to a poor representation of the network structure, due to the heterogeneity in this sub-population.
In our approach, uniform subsampling often leads to good performance; see Sections~\ref{sec:sim} and \ref{sec:app}.
As an alternative, we also propose an adaptive sampling mechanism, which relies on the currently estimated network structure to draw an informative sample of nodes.
This adaptive strategy samples, at iteration $t$, each node $j\in \mathcal{J}_{i0}^\star$ with a probability proportional to
\begin{equation}
	\label{eq:adas}
	g^{-1}\left(\Big[\bmu_i^{(t-1)}\Big]^\intercal\left[\bmu_j^{(t-1)}\right]\right), \quad \bmu_j^{(t-1)} := \E_{q^{(t-1)}(\w_j)}[\w_j],\quad j\in \mathcal{J}_{i0}^\star,
\end{equation}
which corresponds to the current prediction for the probability to observe and edge between node $i$ and node $j$, according to the underlying \textsc{lfm}.
Therefore, we expect that nodes $j\in \mathcal{J}_{i0}^\star$ more similar to node $i$---according to the latent structure specification---are more likely to be sampled and contribute to the update of $q(\w_i;\blambda_{i})$ at iteration $t$.
From a computational perspective, this strategy requires an additional loop over $j\in \mathcal{J}_{i0}^\star$ to compute all the products outlined in \Cref{eq:adas}. 
This operation might increase the computational time in high-dimensional settings, without significantly affecting the storage; see \Cref{sec:sim} for an empirical evaluation.
With the logit link, the proposed adaptive sampling can be used modifying the factor $r_{i0}$ in Equation~\eqref{eq:gradsplit} into $m_{i0} / m_{i0}^\star$,
with 
\begin{equation}
	\label{eq:adawL}
m_{i0} := \sum_{j \in \mathcal{J}_{i0}} \left[1+\exp\left(-\Big[\bmu_i^{(t-1)}\Big]^\intercal\left[\bmu_j^{(t-1)}\right]\right)\right]^{-1}\negthinspace\negthinspace , \quad m_{i0}^\star := \sum_{j \in \mathcal{J}_{i0}^\star} \left[1+\exp\left(-\Big[\bmu_i^{(t-1)}\Big]^\intercal\left[\bmu_j^{(t-1)}\right]\right)\right]^{-1} \negthinspace\negthinspace \negthinspace. \\
\end{equation}
Similarly, for the probit link function we set
\begin{equation}
	\label{eq:adawP}
	m_{i0} := \sum_{j \in \mathcal{J}_{i0}} \Phi\left(\Big[\bmu_i^{(t-1)}\Big]^\intercal\left[\bmu_j^{(t-1)}\right]\right), \quad
m_{i0}^\star := \sum_{j \in \mathcal{J}_{i0}^\star} \Phi\left(\Big[\bmu_i^{(t-1)}\Big]^\intercal\left[\bmu_j^{(t-1)}\right]\right),
\end{equation}
where $\Phi(x)$ denotes the cumulative distribution function of a standard Gaussian evaluated at $x$.
Pseudo code illustrating \textsc{svilf} is reported in \Cref{alg:ssviL,alg:ssviP}, and an R package implementing the methods is available at \url{github.com/emanuelealiverti/svilf} and in the Supplementary Materials.

\begin{algorithm}[!t]
	\caption{\textsc{svilf} for \textsc{lfm} with logit link.} \label{alg:ssviL}
	Initialize $\left\{\blambda_1^{(1)}, \dots, \blambda_n^{(1)}\right\}$ and set step size sequence $\rho_{t}$. \\
 \For( ){{\normalfont{$t=2$ until convergence}}}
 {
	 Sample a permutation $\sigma$ of $\{1,\dots,n\}$ uniformly\\
\For( ){{\normalfont{$i=\sigma(1),\dots,\sigma(n)$}}}
 {
\vspace{7pt}
{\bf [1] Sampling } Construct the random set $\mathcal{J}_i^\star = \mathcal{J}_{i1} \cup \mathcal{J}_{i0}^\star$ using uniform or adaptive subsampling \\
{\bf [2] Local optimization} Compute the locally optimized densities for $z_{ij}, j \in \mathcal{J}_i^\star $ leading to P\`{o}lya-Gamma densities with natural parameters
\begin{eqnarray*}
	\psi_{ij}\left(\blambda_i^{(t-1)}, \blambda_j^{(t-1)}\right) = -0.5 \left[\mbox{vec}\left(\bS_i^{(t-1)}\right)^\intercal\mbox{vec}\left(\bS_j^{(t-1)}\right)\right],
\end{eqnarray*}
with 
\begin{eqnarray*}
	\bS_j^{(t-1)} =  \left(-2\blambda_{j2}^{(t-1)}\right)^{-1} + \left[\left(-2\blambda_{j2}^{(t-1)}\right)^{-1} \blambda_{j1}^{(t-1)}\right]^{\intercal} \left[\left(-2\blambda_{j2}^{(t-1)}\right)^{-1} \blambda_{j1}^{(t-1)}\right].
\end{eqnarray*}
\vspace{7pt}

{\bf [2] Global optimization.} Update the global parameters leveraging stochastic optimization. \\
\vspace{-10pt}
\begin{alignat*}{3}
	\blambda_{i1}^{(t)} = 	(1-\rho_{t}) \blambda_{i1}^{(t-1)} + \rho_{t}\Bigg\{
	\sum_{j \in \mathcal{J}_{i1}} \left[\left(-2\blambda_{j2}^{(t-1)}\right)^{-1} \blambda_{j1}^{(t-1)}\right] \left(y_{ij}-0.5\right)+  \\ r_{i0} \sum_{j \in \mathcal{J}_{i0}^\star}\left[\left(-2\blambda_{j2}^{(t-1)}\right)^{-1} \blambda_{j1}^{(t-1)}\right] \left(y_{ij}-0.5\right) + \I_H\ba_0 \Bigg\} \\
	\blambda_{i2}^{(t)} = 	(1-\rho_{t}) \blambda_{i2}^{(t-1)} - \rho_{t} 0.5\left(\sum_{j \in \mathcal{J}_{i1}} \bS_j^{(t-1)} \cdot \bar{z}_{ij}^{(t-1)} + r_{i0}\sum_{j \in \mathcal{J}_{i0}^\star} \bS_j^{(t-1)} \cdot \bar{z}_{ij}^{(t-1)}  + \I_H\right),
\end{alignat*}
with 
\begin{equation*}
	\bar{z}_{ij}^{(t-1)} = \left[\psi_{ij}\left(\blambda_i^{(t-1)}, \blambda_j^{(t-1)}\right) \right]^{-1} \tanh\left[\psi_{ij}\left(\blambda_i^{(t-1)}, \blambda_j^{(t-1)}\right) \right]
\end{equation*}
and with $r_{i0} = n_{i0}/n^\star_{i0}$ in case of uniform sampling and $r_i = m_{i0}/m^\star_{i0}$ for the adaptive version, as outlined in \Cref{eq:adawL}.
Therefore, the approximating density for $\w_i$ is Gaussian with mean $\bmu_i^{(t)}$  and covariance $\bSigma_i^{(t)}$, where
\begin{equation*}
\bmu_i^{(t)} = \left(-2\blambda_{i2}^{(t)}\right)^{-1} \blambda_{i1}^{(t)}, \quad \bSigma_i^{(t)}= \left(-2\blambda_{i2}^{(t)}\right)^{-1}.
\end{equation*}

}

}
{\bf Output} $q^\star(\W)=\prod_{i=1}^n q^\star(\w_i)$.
\vspace{2pt}
\end{algorithm}

\newpage
		    {\footnotesize
\begin{algorithm}[H]
	\caption{\textsc{svilf} for \textsc{lfm} with probit link.} \label{alg:ssviP}
	Initialize $\left\{\blambda_1^{(1)}, \dots, \blambda_n^{(1)}\right\}$ and set step size sequence $\rho_{t}$. \\
 \For( ){{\normalfont{$t=2$ until convergence}}}
 {
	 Sample a permutation $\sigma$ of $\{1,\dots,n\}$ uniformly\\

 \For( ){{\normalfont{$i=1,\dots,n$}}}
 {
\vspace{7pt}
{\bf [1] Sampling } Construct the random set $\mathcal{J}_i^\star = \mathcal{J}_{i1} \cup \mathcal{J}_{i0}^\star$ using uniform or adaptive subsampling \\
{\bf [2] Local optimization} Compute the locally optimized densities for $z_{ij}, j \in \mathcal{J}_i^\star $ leading to Truncated Normal distributions with natural parameters
\begin{eqnarray*}
	\psi_{ij}\left(\blambda_i^{(t-1)}, \blambda_j^{(t-1)}\right) =\left[\left(-2\blambda_{i2}^{(t-1)}\right)^{-1} \blambda_{i1}^{(t-1)}\right]^{\intercal} \left[\left(-2\blambda_{j2}^{(t-1)}\right)^{-1} \blambda_{j1}^{(t-1)}\right]
\end{eqnarray*}
{\bf [2] Global optimization.} Update the global parameters leveraging a stochastic optimization \\
	 \begin{alignat*}{3}
	 \blambda_{i1}^{(t)} = 	(1-\rho_{t}) \blambda_{i1}^{(t-1)} + \rho_{t}\Bigg\{
	 \sum_{j \in \mathcal{J}_{i1}} \left[\left(-2\blambda_{j2}^{(t-1)}\right)^{-1} \blambda_{j1}^{(t-1)}\right] \tilde{z}_{ij} + \\ \vphantom{m} r_{i0}\sum_{j \in \mathcal{J}_{i0}^\star}\left[\left(-2\blambda_{j2}^{(t-1)}\right)^{-1} \blambda_{j1}^{(t-1)}\right]\tilde{z}_{ij} + \I_H\ba_0 \Bigg\}  \\
	\blambda_{i2}^{(t)} = 	(1-\rho_{t}) \blambda_{i2}^{(t-1)} - \rho_{t} 0.5\left(\sum_{j \in \mathcal{J}_{i1}} \bS_j^{(t-1)}+ r_{i0}\sum_{j \in \mathcal{J}_{i0}^\star} \bS_j^{(t-1)}+ \I_H\right),
\end{alignat*}
\vspace{-3pt}
with 
\vspace{-3pt}
\begin{align*}
	\bS_j^{(t-1)} =  \left(-2\blambda_{j2}^{(t-1)}\right)^{-1} + \left[\left(-2\blambda_{j2}^{(t-1)}\right)^{-1} \blambda_{j1}^{(t-1)}\right]^{\intercal} \left[\left(-2\blambda_{j2}^{(t-1)}\right)^{-1} \blambda_{j1}^{(t-1)}\right]
\end{align*}
\vspace{-3pt}
and 
\vspace{-3pt}
\begin{align*}
	\tilde{z}_{ij}^{(t-1)} &= \psi^{(t-1)}_{ij} \left(\blambda_i^{(t-1)}, \blambda_j^{(t-1)}\right) + \\ & (2y_{ij}-1)\phi\left(\psi_{ij}^{(t-1)} \left(\blambda_i^{(t-1)}, \blambda_j^{(t-1)}\right) \right)\Phi\left[(2y_{ij}-1)\psi_{ij}^{(t-1)} \left(\blambda_i^{(t-1)}, \blambda_j^{(t-1)}\right) \right]^{-1},
\end{align*}
where $\phi(x)$ and $\Phi(x)$ represent the density and the cumulative distribution function of a standard Gaussian evaluated in $x$, respectively, and with $r_{i0} = n_{i0}/n^\star_{i0}$ in case of uniform sampling and $r_i = m_{i0}/m^\star_{i0}$ for the adaptive version, as outlined in \Cref{eq:adawP}.
}
Therefore, the approximating density for $\w_i$ is Gaussian with mean $\bmu_i^{(t)}$  and covariance $\bSigma_i^{(t)}$, where
\begin{equation*}
\bmu_i^{(t)} = \left(-2\blambda_{i2}^{(t)}\right)^{-1} \blambda_{i1}^{(t)}, \quad \bSigma_i^{(t)}= \left(-2\blambda_{i2}^{(t)}\right)^{-1}.
\end{equation*}

}
{\bf Output} $q^\star(\W)=\prod_{i=1}^n q^\star(\w_i)$.
\end{algorithm}
}

\section{Simulation studies}
\label{sec:sim}
\begin{table}[H]
\caption{Simulation studies. Sampling distributions for the three considered scenarios.\vspace{2pt}}
	\label{tab:samplingDist}
	\centering
\begin{tabular}{lcc}
		&  Model              & Sampling distribution \\
		\toprule
		\textsc{s1} & Latent factor model & 
		$\begin{array}{r@{}c@{}}
		&	(y_{ij} \mid \pi_{ij}) \sim \mbox{Ber}(\pi_{ij}), \quad i < j \\ 
		&	\pi_{ij} = [1 + \exp(-\w_i^\intercal \w_j)]^{-1}, \\
		&	\w_i  \sim \mbox{N}_2(\mathbf{0}, 9\, \mathbf{I}_2), \quad i = 1, \dots,n
		    \end{array}$\\
		\midrule
		\textsc{s2} & Latent distance model & 
$\begin{array}{r@{}c@{}}
			& (y_{ij} \mid \pi_{ij}) \sim \mbox{Ber}(\pi_{ij}), \quad i < j \\ 
			& \pi_{ij} = [1 + \exp(-|| \w_i - \w_j||_2)]^{-1}, \w_i \in \mathbb{R}^2 \\
			& \w_i  \sim \mbox{N}_2(\mathbf{0}, \mathbf{I}_2),  \quad i = 1, \dots,n
		    \end{array}$
		\\
		    \midrule
			\textsc{s3} & Stochastic block model & 
$\begin{array}{r@{}c@{}}
			&(y_{ij} \mid \pi_{ij}) \sim \mbox{Ber}(\pi_{ij}), \quad i < j \\ 
			&\pi_{ij} =0.6 \, \mbox{I}\{w_i = w_j\} + 0.2 \, \mbox{I}\{w_i \neq w_j\} \\ 
			&w_i \sim \mbox{Ber}( 0.5),  \quad i = 1, \dots,n
		    \end{array}$\\
			\bottomrule
	\end{tabular}
\end{table}

We conduct a simulation study to evaluate the performances of the proposed algorithms.
To compare our methods with a similar alternative---in terms of variational approximation under the same statistical model---we used the Automated Differentiation Variational Inference algorithm \citep[\textsc{advi},][]{kucukelbir:2015} to approximate the posterior distribution of the \textsc{lfm} specified in Equation~\ref{eq:lfm}. 
This algorithm is available in the software \textsc{stan}~\citep{stan}, and it relies on a fully-factorized Gaussian approximation whose parameters are obtained via stochastic optimization, leveraging an adaptive step-size sequence; see \citet{kucukelbir:2015} for more details.
We also compare \textsc{svilf} with a Latent distance model \citep{hoff:2002} estimated via Variational-\textsc{em} \citep[\textsc{vblpcm},][]{vblcp}.

In this section, we focus on the  logit link for the \textsc{lfm}; results for the probit link are reported in the Supplementary Materials.
Computational performances are evaluated in terms of memory usage, elapsed time and goodness of fit.  Memory is measured in \textsc{mb} of used \textsc{ram}, while elapsed time is evaluated in seconds to run the \textsc{vb} routine until convergence.
For \textsc{svilf} and \textsc{vblpcm}, we use as convergence criterion the mean squared difference between consecutive parameter values below $10^{-5}$; for \textsc{advi}, we follow the default implementation, relying on a median \textsc{elbo} difference below $10^{-2}$.
Predictions are evaluated in terms of adequacy in recovering the probability of observing an edge, according to the Area Under the Roc curve (\textsc{auc}).

The simulations focus on three different data generating processes, summarized in  \Cref{tab:samplingDist}. Networks in the first scenario (\textsc{s1}) are generated according to a \textsc{lfm} model, with $2$ latent factors randomly generated from multivariate Gaussian with diagonal covariance and standard deviation equal to $3$.
In the second simulation scenario (\textsc{s2}), data are generated from a latent distance model \citep{hoff:2002} having $2$-dimensional latent positions, generated from standard Gaussian distributions, while in the third and last scenario (\textsc{s3}), data are generated from a stochastic block-model with $2$ latent groups, equal weights, within-group probability of connection equal to $0.6$ and between-group probability equal to $0.2$.

We focus on networks with a number of nodes ${n \in \{100,200,300,500,1000,2000,3000\}}$, and replicate the data generation process and posterior estimation with $100$ different random seeds.
We fix the number of latent factors and latent coordinates to $H=4$, set $\ba_0 = \mbox{logit}(2\sum_{i < j} y_{ij}/n(n-1))\cdot \I_H$ to account for the overall network sparsity, and set the \textsc{svilf} step-size parameters $\alpha=1, \beta=0.75, \gamma = 2$. 
In the Supplementary Materials, we extend the simulations to different values of latent factors $H$ and of the parameter $\gamma$.
Posterior inference for \textsc{advi} and \textsc{vblpcm} is performed with default parameters configuration. 
\begin{table}[t]
\caption{Simulation studies. Average \textsc{ram} usage (in \textsc{mb}) across $100$ simulation replicates,  measured with the \texttt{R} function \texttt{gc}. Standard deviations are not reported since memory allocation is essentially constant across simulation replicates.
	}
\centering
\centering
\begin{tabular}{llrrrrrrr}
& $n$ & 100 & 200 & 300 & 500 & 1000 & 2000 & 3000 \\ 
\toprule
\textsc{s1}           & \textsc{svilf}      & 101 & 101 & 101 & 101 & 103 & 108  & 116 \\
	              & \textsc{svilf (ada)}& 101 & 101 & 101 & 101 & 103 & 108  & 116 \\
	              & \textsc{advi}       & 112 & 115 & 119 & 128 & 156 & 227  & 308 \\
	              & \textsc{vblpcm}     & 117 & 122 & 131 & 158 & 285 & 792  & 1635 \\
		      \midrule                                                             
\textsc{s2}           & \textsc{svilf}      & 101 & 101 & 101 & 101 & 103 & 109  & 119 \\
                      & \textsc{svilf (ada)}& 101 & 101 & 101 & 101 & 103 & 109  & 119 \\
                      & \textsc{advi}       & 112 & 115 & 119 & 128 & 156 & 226  & 308 \\
                      & \textsc{vblpcm}     & 117 & 123 & 133 & 164 & 310 & 891  & 1861 \\
		      \midrule                                                             
\textsc{s3}           & \textsc{svilf}      & 101 & 101 & 101 & 102 & 104 & 113  & 129 \\
                      & \textsc{svilf (ada)}& 101 & 101 & 101 & 102 & 104 & 113  & 129 \\
                      & \textsc{advi}       & 112 & 115 & 119 & 128 & 156 & 224  & 308 \\
                      & \textsc{vblpcm}     & 118 & 127 & 141 & 186 & 395 & 1234 & 2631 \\
		      \bottomrule
\end{tabular}

\label{sim:ram}
\end{table}
	The \textsc{advi} algorithm often failed to converge with random initialization of the parameters. Instead,  convergence criteria are met when the parameters are initialized  from the eigenvalues of the adjacency matrix. For the \textsc{svilf} algorithms parameters are initialized randomly, and in most settings convergence is reached in less than $50$ iterations. We found no difference when initialing the parameters from the eigenvalues of the adjacency matrix, and since  eigen-decomposition can be demanding in high-dimensional settings, we recommend  random initialization as a default option for the \textsc{svilf}. For \textsc{vblpcm} we used the build-in function \texttt{vblpcmstart} that generates sensible parameters' initialization.

\Cref{sim:ram} reports the average memory usage for posterior inference in all the considered simulation scenarios, across $100$ simulation replicates. 
When the number of nodes is in the order of few hundred, the considered algorithms require a similar amounts of \textsc{ram}. 
In contrast, when the number of nodes is in the order of few thousand, the proposed \textsc{svilf} algorithms requires significantly less memory in all the scenarios.
For example, in the case $n=3000$, \textsc{advi} and \textsc{vblpcm} require approximately $3$ and $20$ times as much memory as \textsc{svilf}; this factor is increasing with $n$, showing the efficiency of the \textsc{svilf} algorithms for high-dimensional networks.
As expected, the two sampling schemes implemented under the \textsc{svilf} algorithm perform identically in terms of memory usage, since no additional storage capacity is needed.

\begin{table}[t]
\centering
\caption{Simulation studies. Average elapsed time (standard deviation) in seconds across $100$ simulation replicates.}
\resizebox{\textwidth}{!}{
\begin{tabular}{llrrrrrrrr}
& $n$ & 100 & 200 & 300 & 500 & 1000 & 2000 & 3000 \\  
\toprule
\textsc{s1}            & \textsc{svilf}       & 1.2 (0.1) & 2.0 (0.3)  & 4.2 (0.8)  & 14.8 (1.7)  & 112 (12)  & 893 (78)    & 3362 (337) \\   
	               & \textsc{svilf (ada)} & 2.5 (0.2) & 7.6 (0.9)  & 15.2 (1.6) & 37.6 (4.6)  & 188 (22)  & 1212 (126)  & 4270 (542) \\   
	               & \textsc{advi}        & 7.4 (1.2) & 27.8 (2.5) & 62.4 (7.0) & 176.3 (14.7)& 711 (49)  & 2815 (229)  & 6543 (451) \\   
	               & \textsc{vblpcm}      & 5.4 (0.5) & 19.5 (1.4) & 48.5 (8.8) & 159.9 (34.7)& 738 (207) & 3761 (901)  & 10247 (2334) \\ 
\midrule                                                                                                                                              
\textsc{s2}           & \textsc{svilf}       & 1.2 (0.1) & 1.8 (0.2)  & 2.9 (0.4)  & 7.2 (1.1)    & 37 (4)    & 256 (42)    & 1017 (206) \\   
                      & \textsc{svilf (ada)} & 2.5 (0.2) & 7.7 (0.5)  & 15.9 (1.1) & 47.4 (3.7)   & 244 (19)  & 1445 (137)  & 4698 (629) \\   
                      & \textsc{advi}        & 6.8 (1.6) & 25.1 (2.8) & 58.8 (9.7) & 165.8 (16.2) & 658 (46)  & 2578 (177)  & 6043 (427) \\   
                      & \textsc{vblpcm}      & 5.0 (0.5) & 18.9 (1.5) & 47.4 (9.0) & 105.1 (11.3) & 744 (210) & 3977 (1013) & 11711 (2622) \\ 
\midrule                                                                                                                                              
\textsc{s3}           & \textsc{svilf}       & 1.2 (0.1) & 1.9 (0.3)  & 3.3 (0.3)  & 9.0 (1.0)    & 48 (4)    & 332 (33)    & 1306 (90) \\    
                      & \textsc{svilf (ada)} & 2.9 (0.2) & 9.1 (0.6)  & 19.8 (1.2) & 56.1 (3.6)   & 276 (15)  & 1622 (167)  & 5785 (597) \\   
                      & \textsc{advi}        & 6.7 (0.6) & 25.3 (2.2) & 57.1 (5.3) & 163.2 (19.1) & 647 (40)  & 2542 (154)  & 5926 (375) \\   
                      & \textsc{vblpcm}      & 4.8 (0.6) & 18.5 (1.5) & 48.3 (9.3) & 163.1 (37.8) & 960 (174) & 4586 (935)  & 14848 (2878) \\ 
  \bottomrule                                                                                      
\end{tabular}                                                                                      

}
\label{sim:tabtime}
\end{table}

\Cref{sim:tabtime} compares the elapsed time required by the different algorithms to reach convergence.
Results suggest that \textsc{svilf} with uniform subsampling provides the fastest routine in all the considered settings, with an elapsed $2$ to $15$ times faster than \textsc{advi} and \textsc{vblpcm}.
For example, with $n=2000$ and data generated under a latent-distance model (\textsc{s2}), \textsc{svilf} is $13$ times faster than the competitors, requiring in $4$ minutes on average versus $42$ of \textsc{advi} and $66$ of \textsc{vblpcm}. 
As expected, the adaptive sampling scheme affects the execution time of the \textsc{svilf} algorithm, because of the additional loop of order $\mathcal{O}(n_{i0})$ to compute the sampling weights at each iteration.
However,  even if this adaptive strategy is more expensive than the standard \textsc{svilf}, it still provides significant computational advantages over \textsc{advi} and \textsc{vblpcm}, reducing the elapsed time and memory usage.

\begin{table}[t]
\centering
\caption{Simulation studies. Average (standard deviation) \textsc{auc} in percentage values across $100$ simulation replicates.}

\begin{tabular}{llrrrrrrrr}
& $n$ & 100 & 200 & 300 & 500 & 1000 & 2000 & 3000 \\  
\toprule
\textsc{s1}            & \textsc{svilf}      & 86 (1.6) & 86 (1.3) & 86 (0.9) & 85 (0.7) & 85 (0.5) & 85 (0.4) & 85 (0.4) \\
	               & \textsc{svilf (ada)}& 86 (1.3) & 85 (1.2) & 85 (1.0) & 85 (0.9) & 84 (0.6) & 84 (0.4) & 84 (0.4) \\
	               & \textsc{advi}       & 85 (2.1) & 86 (1.2) & 86 (0.9) & 86 (0.7) & 86 (0.5) & 86 (0.4) & 86 (0.3) \\
	               & \textsc{vblpcm}     & 87 (1.3) & 88 (1.0) & 87 (0.9) & 87 (0.9) & 84 (3.3) & 82 (4.1) & 76 (7.2) \\
\midrule                                                                                                                     
\textsc{s2}           & \textsc{svilf}       & 72 (0.9) & 68 (0.7) & 66 (0.5) & 64 (0.4) & 63 (0.4) & 62 (0.3) & 62 (0.3) \\
                      & \textsc{svilf (ada)} & 73 (0.7) & 69 (0.5) & 67 (0.5) & 66 (0.4) & 65 (0.4) & 65 (0.3) & 65 (0.2) \\
                      & \textsc{advi}        & 66 (2.8) & 65 (1.1) & 66 (1.0) & 66 (0.6) & 64 (0.5) & 63 (0.3) & 62 (0.2) \\
                      & \textsc{vblpcm}      & 78 (0.8) & 73 (0.6) & 71 (0.5) & 69 (0.5) & 67 (0.6) & 67 (0.4) & 61 (2.0) \\
\midrule                                                                                                                     
\textsc{s3}           & \textsc{svilf}       & 80 (0.6) & 77 (0.4) & 76 (0.2) & 75 (0.2) & 74 (0.1) & 73 (0.1) & 72 (0.1) \\
                      & \textsc{svilf (ada)} & 80 (0.6) & 77 (0.3) & 76 (0.2) & 75 (0.1) & 74 (0.1) & 73 (0.1) & 72 (0.1) \\
                      & \textsc{advi}        & 75 (1.5) & 74 (0.4) & 73 (0.3) & 73 (0.2) & 72 (0.2) & 72 (0.1) & 72 (0.1) \\
                      & \textsc{vblpcm}      & 80 (0.6) & 76 (0.3) & 75 (0.2) & 73 (0.2) & 72 (0.1) & 71 (0.2) & 64 (6.7) \\
  \bottomrule                                                                                      
\end{tabular}

\label{sim:auc}
\end{table}

In order to evaluate the quality of predictions, \Cref{sim:auc} compares the goodness of fit of the posterior predicted probabilities in terms of \textsc{auc}, reported in percentage values.
The considered approaches share similar performances in most simulation scenarios. 
In the second Scenario, \textsc{vblpcm} is the correctly specified model and it performs better than other methods on average. However, the performance gap is decreasing with the size of $n$, and when $n=3000$ the \textsc{auc} are not significantly different.
Therefore, the main advantage of \textsc{svilf} is to provide massive savings in terms of computational time and memory usage, achieving accurate performance in reasonable time also in  high-dimensional settings. 
In the Supplementary Materials, we include additional simulations to illustrate the performance of \textsc{svilf} with different specification of the number of latent factors $H$ and the parameter $\gamma$, suggesting that the performance can be improved increasing $H$ without affecting the computational performance.

\section{Application}
\label{sec:app}

We analyze three high-dimensional datasets provided by \citet{rozemberczki:2019} and freely available on the repository \url{http://snap.stanford.edu/data/index.html}.
The main aim of this section is to show how \textsc{svilf} algorithms provide accurate representations of high-dimensional network data in terms of link prediction.

The first dataset involves mutual interconnections among Facebook public pages related to politicians, governmental organizations, television shows and companies; the resulting network consists of $n=22700$ nodes corresponding to verified Facebook pages, while edges represent the mutual ``likes'' between these pages.   
The second dataset includes the mutual following relationships among $n=37700$ users of the popular software developing platform GitHub. 
The third dataset includes $n=11631$ articles on crocodiles from the English version of Wikipedia; an edge indicates mutual links between articles.
In all the examples, we estimate an \textsc{lfm} using \textsc{svilf} algorithms with the same settings as in the simulations studies, increasing $\gamma = 3$ to handle the massive sparsity of the networks (the average density is $0.001$). 
We did not compare \textsc{svilf} with the competitors as the number of nodes is excessively large.

\begin{table}[b!]
\caption{Application. \textsc{auc} for the proposed methods.}
\centering
\begin{tabular}{rlrr}
& & \textsc{logit} & \textsc{probit} \\
\midrule
	Facebook                   & \textsc{svilf (ada)} & 0.855 & 0.835 \\
	                              & \textsc{svilf}       & 0.678 & 0.798 \\
	GitHub                  & \textsc{svilf (ada)} & 0.791 & 0.867 \\
	                              & \textsc{svilf}       & 0.874 & 0.809 \\
			Wikipedia & \textsc{svilf (ada)} & 0.757 & 0.950 \\
	                              & \textsc{svilf}       & 0.867 & 0.940 \\
		      \bottomrule
\end{tabular}
\label{tab:results}
\end{table}

\begin{figure}[t]
	\centering
	\includegraphics[width=\textwidth]{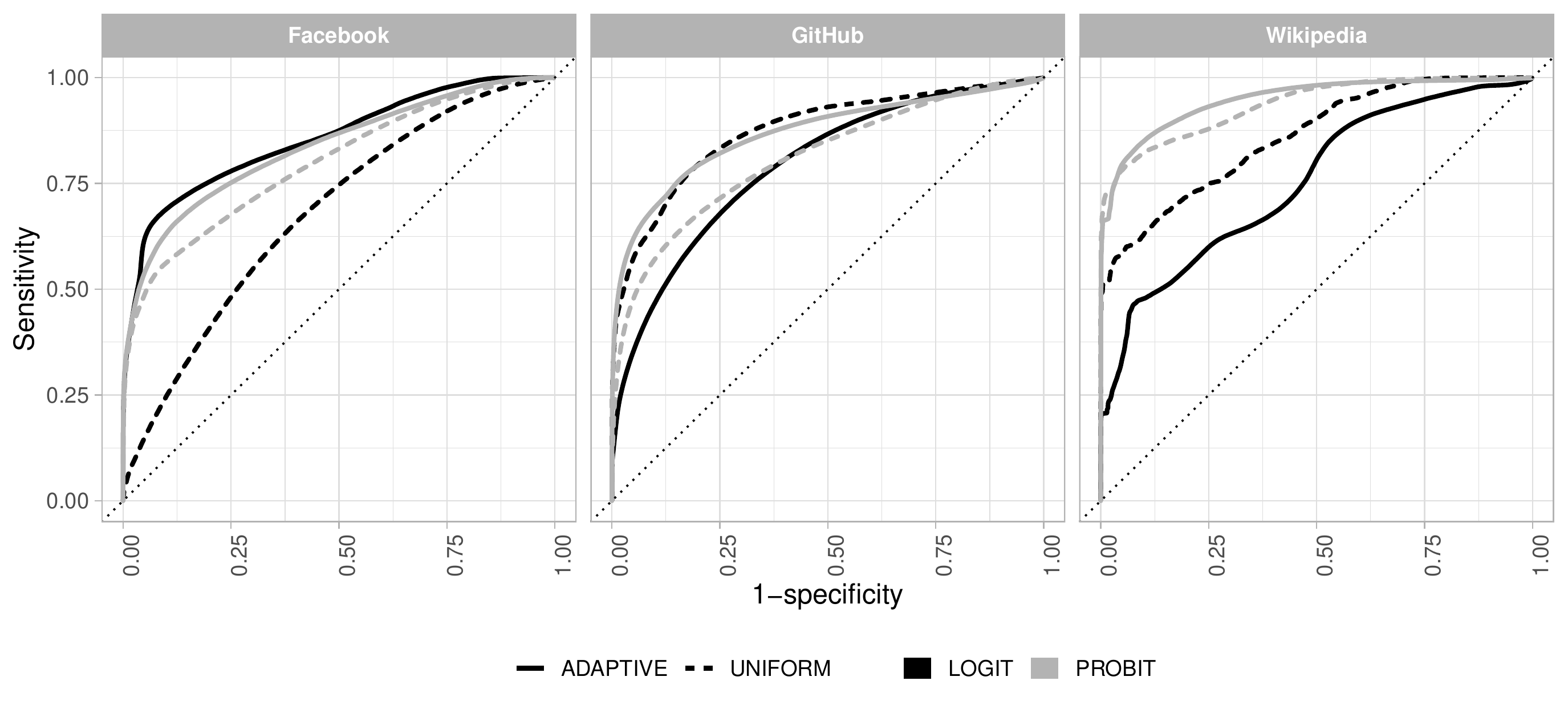}
	\caption{Application. Smoothed \textsc{roc} curves for the proposed approaches over the three examples.}
	\label{fig:roc}
\end{figure}

\Cref{tab:results} and \Cref{fig:roc} illustrate, respectively, the \textsc{auc} and the \textsc{roc} curves  for the considered approaches. 
Results suggest a satisfactory performance for the proposed methods, with an \textsc{auc} above $0.75$ in most settings. 
Coherently with the simulations, our empirical findings do not indicate that the adaptive subsampling is systematically preferable over uniform sampling. 
Results from \Cref{tab:results} suggest that the best-performing approach is different across applications; for example, uniform subsampling achieves lower \textsc{auc} than the adaptive approach in the Facebook example, and higher in the GitHub dataset.
Such differences might be related to the structure of network, and it will be of interest in future developments to investigate under which network structure one particular sampling approach is preferable.

\section{Discussion}
Motivated by the abundance of large network-valued data, we have proposed a novel computational approach for approximate posterior inference under the latent factor model for networks. Specifically, we developed a stochastic Variational Bayes routine that explicitly leverages the sparsity of high-dimensional networks to perform efficient computation.
The empirical evaluations suggest that the proposed algorithms lead to significant benefits in terms of computational efficiency, without affecting accuracy in modeling the network connectivity structure.

Although the \textsc{lfm} is routinely used to model undirected binary networks, several applications involve more complicated structures.
Some examples includes multiple networks, presence of weighted edges or the desire to include additional covariates into the analysis.
The \textsc{svilf} algorithms for the \textsc{lfm} can be directly extended to such settings considering different conditional distributions for the elements $y_{ij}$ (e.g., Poisson or Gaussian), and allowing the probability of observing an edge to change according to some edge-specific covariates.
Additionally, the subsampling strategy proposed in Section~\ref{sec:svilf} can be potentially adapted to approximate the gradients in algorithms for latent distance models \citep[e.g.,][]{vblcp,gollini:2016}.  However, the implementation requires ad-hoc steps, since these models do not fall under the same conditionally conjugate structure used by \textsc{svilf}.  

\section{Acknowledgements}
The work of Emanuele Aliverti was partially funded by \textsc{miur-prin} 2017 project \textsc{2017br-jxs}, as well as grant  grant \textsc{bird}-188753/18 of the University of Padova, Italy. The authors would like to thank Bruno Scarpa, Peter Hoff and Daniele Durante for their comments and suggestions on the main idea of this work.
This research used the \textsc{vera} cluster system at University Ca’ Foscari Venezia.

\bibliographystyle{apalike}
\bibliography{AR_svilf}
\end{document}


\def\spacingset#1{\renewcommand{\baselinestretch}%
{#1}\small\normalsize} \spacingset{1}


\title{\bf Stratified stochastic variational inference for high-dimensional network factor model \\[1cm] Supplementary materials}
\author{Emanuele Aliverti and Massimiliano Russo}
\vspace{-3cm}
\maketitle

\spacingset{1.5} 

\section{Simulations with different latent dimension $H$ and parameter $\gamma$}
We evaluate the performance of the \textsc{svilf} algorithm described in Section~{4} of \citet{aliverti2022:svilf} with all the combination $(H,\gamma) \in \{2, \dots, 10\} \times \{1,\dots5\}$. The simulations focus on the same data generating processes outlined in Section~4 and Table~1 of the manuscript. 

\begin{sidewaysfigure}
\centering
\includegraphics[width = \textwidth]{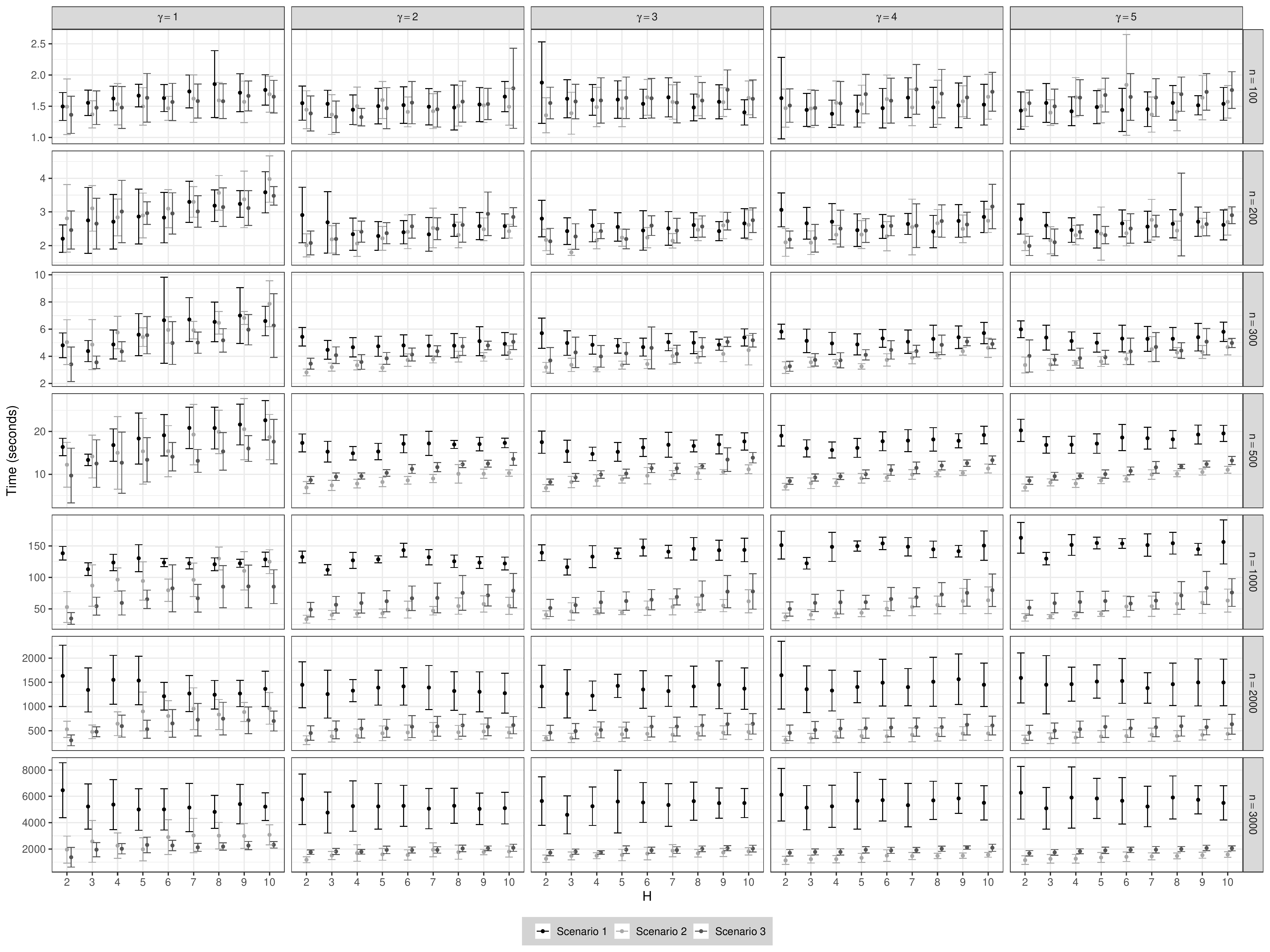}
\caption{Average elapsed time (in seconds) for \textsc{svilf} across $10$ replicates of the simulations described in Section~{4} with $(H,\gamma) \in \{2, \dots, 10\} \times \{1,\dots5\}$. The error-bars represent the average elapsed time plus/minus two standard deviations.}
\label{fig:htime}
\end{sidewaysfigure}

\begin{sidewaysfigure}
\centering
\includegraphics[width = \textwidth]{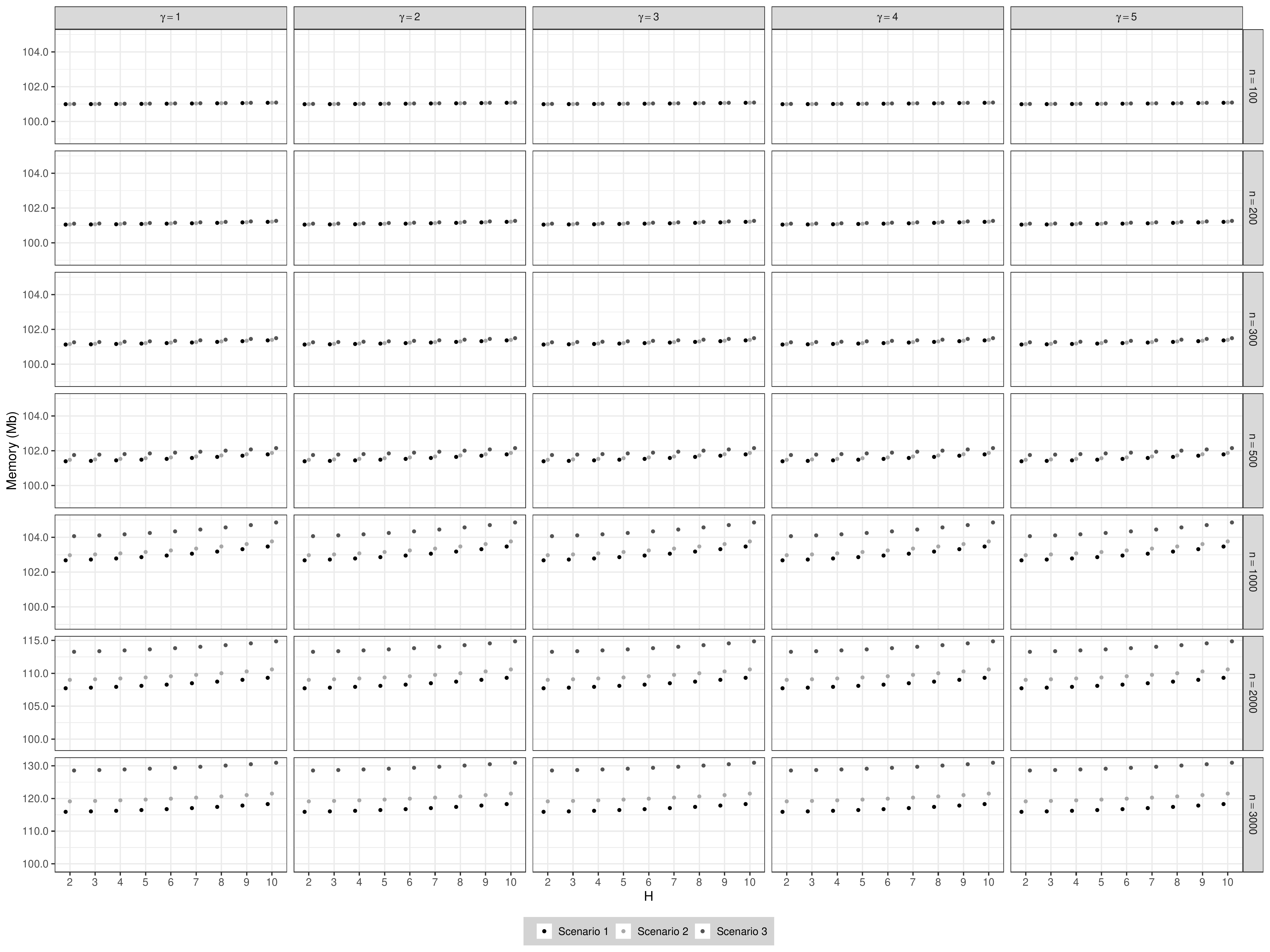}
\caption{Average \textsc{ram} usage (in \textsc{mb}) for \textsc{svilf} across $10$  replicates of the simulations described in Section~{4} with $(H,\gamma) \in \{2, \dots, 10\} \times \{1,\dots5\}$. Memory usage is  measured with the \texttt{R} function \texttt{gc}. Variability is not reported since memory allocation is essentially constant across simulation replicates.}
\label{fig:hmem}
\end{sidewaysfigure}

\begin{sidewaysfigure}
\centering
\includegraphics[width = \textwidth]{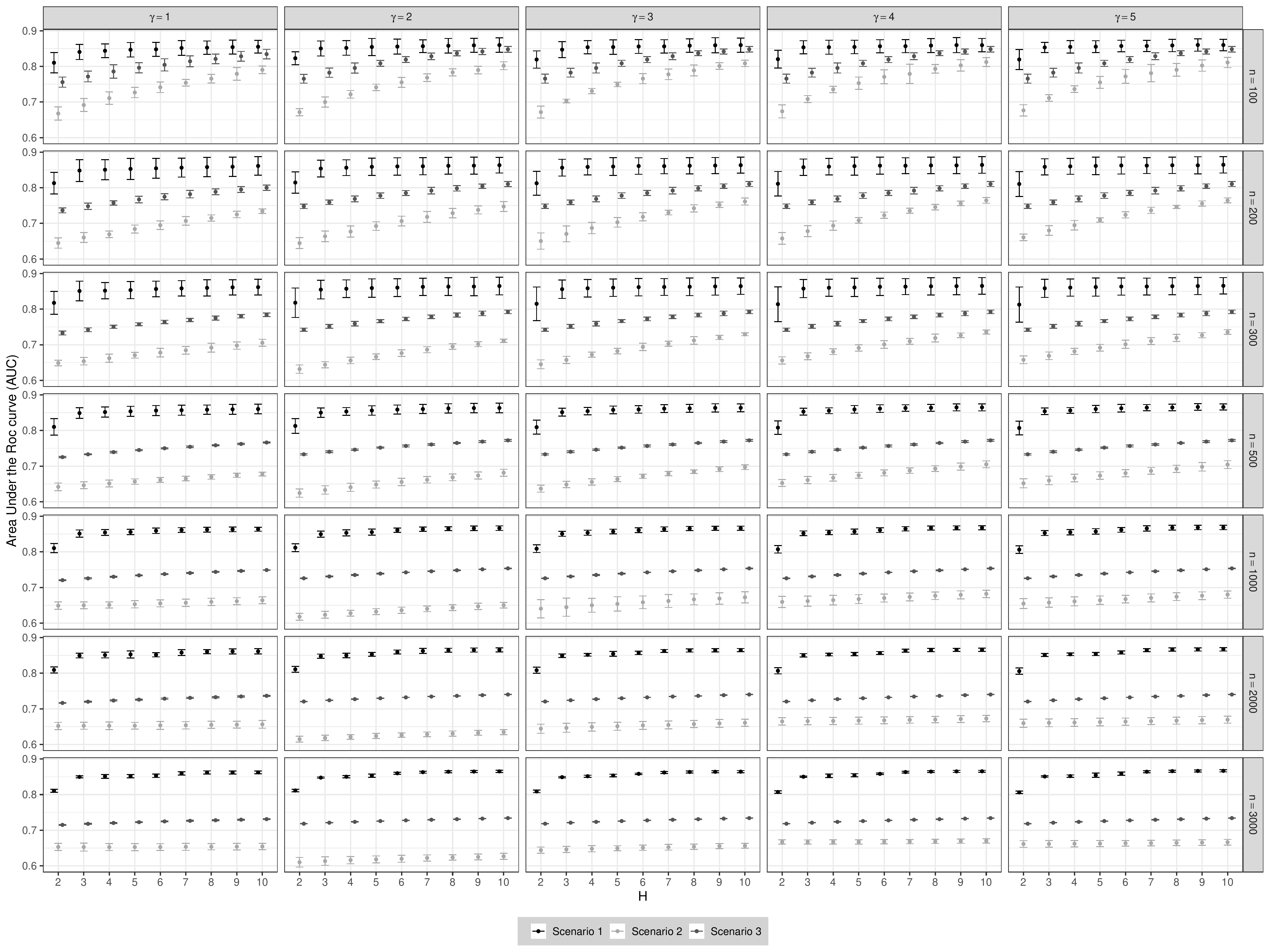}

\caption{Average \textsc{auc} for \textsc{svilf} across $10$  replicates of the simulations described in Section~{4} with $(H,\gamma) \in \{2, \dots, 10\} \times \{1,\dots5\}$. The error-bars represent the average \textsc{auc} plus/minus two standard deviations.}
\label{fig:hauc}
\end{sidewaysfigure}

\Cref{fig:htime} reports the elapsed time for \textsc{svilf}. The dimension of the latent space and the magnitude of $\gamma$ do not affect the performance of the algorithm in terms of timing. For a fixed $n$ the elapsed time is essentially constant for all the  considered $(H,\gamma)$ values. This is expected since, as outlined in Section~4 of the manuscript, the \textsc{svilf} computational cost is dominated by $n$.
\Cref{fig:hmem} shows the memory usage, suggesting that also this quantity is constant for different values of $(H,\gamma)$. The highest inflation in terms of memory usage is observed for  $n = 3000$, where the used \textsc{ram} ranges from $128$ \textsc{mb} with $H=2$ to $132$ \textsc{Mb} with $H=10$, irrespectively of the specific value of $\gamma$.
Lastly, \Cref{fig:hauc} depicts the values for the \textsc{auc}.
In Scenario 1, the \textsc{lfm} is the correctly specified model, and we do not observe performance improvements for $H \ge 3$. In Scenarios 2 and 3 we observe a moderate increasing trend in $H$, with performance that improves for larger number of latent factors. Results are consistent for different values of $\gamma$, suggesting that this parameter does not influence much model performance when selected in a reasonable range.

\section{Simulation studies with probit link}
\label{sec:probitsim}

\begin{table}[h!]
\center
\caption{Simulation studies. Average \textsc{ram} usage (in \textsc{mb})  measured with the \texttt{R} function \texttt{gc} across $100$  replicates of the simulation study. Standard deviations are not reported since memory allocation is essentially constant across simulation replicates.}
\label{tab:1}
\begin{tabular}{llrrrrrrr}
& $n$ & 100 & 200 & 300 & 500 & 1000 & 2000 & 3000 \\ 
  \toprule
  \textsc{s1}  & \textsc{svilf}       & 101 & 101 & 101 & 101 & 103 & 108 & 116 \\
	       & \textsc{svilf (ada)} & 101 & 101 & 101 & 101 & 102 & 108 & 116 \\
  \midrule
  \textsc{s2}  & \textsc{svilf}       & 101 & 101 & 101 & 102 & 103 & 109 & 119 \\
               & \textsc{svilf (ada)} & 101 & 101 & 101 & 102 & 103 & 109 & 119 \\
  \midrule
  \textsc{s3}  & \textsc{svilf}       & 101 & 101 & 101 & 102 & 104 & 113 & 129 \\
               & \textsc{svilf (ada)} & 101 & 101 & 101 & 102 & 104 & 113 & 129 \\
   \bottomrule
\end{tabular}
\end{table}

\begin{table}[h!]
\caption{Simulation studies. Average elapsed time (standard deviation) in seconds across $100$  replicates of the simulation study.}
\resizebox{\textwidth}{!}{
\begin{tabular}{lllllllll}
\label{tab:2}
& $n$ & 100 & 200 & 300 & 500 & 1000 & 2000 & 3000 \\ 
  \toprule
  \textsc{s1} & \textsc{svilf}       & 1.8 (0.5) & 2.1 (0.6)  & 2.8 (0.8) & 5.8 (1.3)   & 36.1 (6.3)   & 397 (71)   & 1460 (188) \\
              & \textsc{svilf (ada)} & 1.9 (0.4) & 3.3 (0.9)  & 5.7 (1.1) & 17.1 (2.3)  & 85.3 (9.0)   & 575 (85)   & 1857 (257) \\
  \midrule
  \textsc{s2} & \textsc{svilf}       & 1.7 (0.4) & 2.1 (0.50) & 2.8 (0.7) & 6.1 (1.2)   & 31.6 (5.1)   & 321 (46)   & 1203 (137) \\
              & \textsc{svilf (ada)} & 2.0 (0.5) & 3.8 (0.58) & 6.7 (1.2) & 17.0 (1.60) & 85.2 (8.4)   & 619 (87)   & 1862 (228) \\
  \midrule
  \textsc{s3} & \textsc{svilf}       & 1.7 (0.5) & 2.4 (0.6)  & 3.9 (1.0) & 9.2 (1.3)   & 55 (9.2)     & 554 (64)   & 1829 (155) \\
              & \textsc{svilf (ada)} & 2.2 (0.5) & 4.7 (0.7)  & 9.6 (0.9) & 26.2 (2.3)  & 152.4 (17.3) & 1326 (178) & 4270 (430) \\
   \bottomrule
\end{tabular}}
\label{tab:3}
\end{table}

\begin{table}[h!]
\caption{Simulation studies. Average (standard deviation) \textsc{auc} in percentage values across $100$ replicates of the simulation study.}
\resizebox{\textwidth}{!}{
\begin{tabular}{lllllllll}
& $n$ & 100 & 200 & 300 & 500 & 1000 & 2000 & 3000 \\ 
  \toprule
 \textsc{s1}   & \textsc{svilf}       & 85 (1.7) & 85 (1.2) & 85 (0.9) & 85 (0.7) & 85 (0.5) & 85 (0.3) & 85 (0.3) \\
               & \textsc{svilf (ada)} & 86 (1.6) & 86 (1.2) & 86 (0.9) & 86 (0.7) & 86 (0.5) & 86 (0.3) & 86 (0.3) \\
  \midrule
 \textsc{s2}   & \textsc{svilf}       & 69 (1.2) & 66 (0.8) & 64 (0.6) & 63 (0.4) & 62 (0.5) & 61 (0.4) & 61 (0.3) \\
               & \textsc{svilf (ada)} & 72 (1.2) & 69 (0.6) & 67 (0.6) & 66 (0.4) & 65 (0.4) & 65 (0.3) & 65 (0.3) \\
  \midrule
 \textsc{s3}   & \textsc{svilf}       & 78 (0.8) & 76 (0.4) & 75 (0.3) & 74 (0.2) & 73 (0.1) & 72 (0.1) & 72 (0.1) \\
               & \textsc{svilf (ada)} & 79 (0.7) & 77 (0.4) & 76 (0.2) & 75 (0.1) & 74 (0.1) & 73 (0.1) & 72 (0.1) \\
   \bottomrule
\end{tabular}}
\label{tab:3}
\end{table}

Results from the Section~4 of \cite{aliverti2022:svilf} are reproduced relying on a \textsc{lfm} with probit link function, implemented within the \texttt{R} package \texttt{svilf} available at \url{https://github.com/emanuelealiverti/svilf}. Tables~\ref{tab:1}-\ref{tab:3} illustrate the average memory, elapsed time and \textsc{auc}, confirming a modest use of computational resources also with large networks, as well as satisfactory performance in terms of \textsc{auc} (\Cref{tab:3}). The results in \Cref{tab:1,tab:2,tab:3} for the probit link function are  comparable with the logit link ones reported in Section~{4} of the paper (Tables 2 to 4)

\section{CAVI algorithm for \textsc{lfm} with probit link}
\Cref{alg:caviP} illustrates the pseudo-code for the \textsc{cavi} algorithm under a \textsc{lfm} with probit link function. 
Leveraging the conditionally-conjugate representation outlined in the article, the implementation is similar with Algorithm 1 of \citet{aliverti2022:svilf}.

\begin{algorithm}[H]
	\small
	\caption{\textsc{cavi} for \textsc{lfm} with probit link.} \label{alg:caviP}
	Initialize $\left\{\bS_1^{(1)}, \dots, \bS_n^{(1)}\right\}$ and  $\left\{\bmu_1^{(1)}, \dots, \bmu_n^{(1)}\right\}$. \\
 \For( ){{\normalfont{$t=2$ until convergence}}}
 {
\vspace{7pt}
	 {\bf [1]} $q^{(t)}(\w_i)$ is a density of a $\mbox{N}_H(\bmu_i^{(t)}, \bSigma_i^{(t)})$ with $\bmu_i^{(t)}=[-2\blambda^{(t)}_{i2}]^{-1} \blambda^{(t)}_{i1}$, $\bSigma_i^{(t)}=[-2\blambda^{(t)}_{i2}]^{-1}$ and
\begin{eqnarray*}
	\blambda^{(t)}_{i1}= \sum_{j\neq i}\bmu^{(t-1)}_j\tilde{z}^{(t-1)}_{ij} + \I_H\ba_0, \quad \blambda^{(t)}_{i2}= -0.5\left(\sum_{j\neq i}\left[ \bSigma_j^{(t-1)} + \left(\bmu_j^{(t-1)}\right)\left(\bmu_j^{(t-1)}\right)^\intercal\right]  + \I_H\right),
\end{eqnarray*}
for $i =1, \dots, n$. In the above expression, 
$$\tilde{z}_{ij}^{(t-1)} = \gamma^{(t-1)}_{ij} + (2y_{ij}-1)\phi(\gamma_{ij}^{(t-1)})\Phi[(2y_{ij}-1)\gamma_{ij}^{(t-1)}]^{-1},\quad \gamma_{ij}^{(t-1)} = \left(\bmu_i^{(t-1)}\right)^\intercal\left(\bmu_j^{(t-1)}\right)$$
and $\phi(x)$ and $\Phi(x)$ represent the density and the cumulative distribution function of a standard Gaussian evaluated in $x$.\\
\vspace{7pt}
{\bf [2]} $q^{(t)}(z_{ij})$, is the density of a $\textsc{TN}[\gamma^{(t)}_{ij}, 1, (-\infty,0)]$ if $y_{ij} = 0$ and $\textsc{TN}[\gamma^{(t)}_{ij}, 1, (0,+\infty)]$ if $y_{ij} = 1$,
for $i=2, \dots, n$ and $j=1,\dots,i-1$.
}
{\bf Output} $q^\star(\W,\z)=\prod_{i=1}^n q^\star(\w_i)\prod_{i=2}^n \prod_{j=1}^{n-1}q^{\star}(z_{ij})$. 
\vspace{2pt}
\end{algorithm}

\bibliographystyle{apalike}
\bibliography{AR_svilf}